\documentclass[conference]{IEEEtran}
\IEEEoverridecommandlockouts
\usepackage{cite}
\usepackage{amsmath,amssymb,amsfonts}
\usepackage{algorithmic}
\usepackage{graphicx}
\usepackage{textcomp}
\usepackage{xcolor}
\def\BibTeX{{\rm B\kern-.05em{\sc i\kern-.025em b}\kern-.08em
    T\kern-.1667em\lower.7ex\hbox{E}\kern-.125emX}}
\begin{document}

\title{Equilibrium Formulation of a 3-DOF Compliant Mechanism using Sylvester's Dialytic Method of Elimination
}

\author{\IEEEauthorblockN{1\textsuperscript{st} Mustafa M. Mustafa}
\IEEEauthorblockA{\textit{Department of Electrical Engineering} \\
\textit{Salahaddin University-Erbil}\\
Erbil, Kurdistan Region, Iraq \\
mustafa.atrushi@su.edu.krd}
\and
\IEEEauthorblockN{2\textsuperscript{nd} Carl D. Crane}
\IEEEauthorblockA{\textit{Department of Mechanical and Aerospace Engineering} \\
\textit{University of Florida}\\
Gainesville, Florida, USA \\
ccrane@ufl.edu}
\and
\IEEEauthorblockN{3\textsuperscript{rd} Ibrahim Hamarash}
\IEEEauthorblockA{\textit{Department of Computer Science and Engineering} \\
\textit{University of Kurdistan Hewler}\\
Erbil, Kurdistan Region, Iraq \\
ibrahim.hamad@ukh.edu.krd}
}

\maketitle

\begin{abstract}
This paper studies the equilibrium formulation of a three degree of freedom planar compliant platform mechanism, which is in contact with a solid body in its environment. The mechanism includes two platforms, which are connected in parallel by three linear springs. The capability of deformation by manipulating both platforms exceptionally complicates the problem. The analysis aims to determine all equilibrium configurations for two different cases: \MakeUppercase{first case} all three springs have zero free lengths and \MakeUppercase{second case} only two of the springs have zero free lengths. The proposed procedure calculates the pose of the top platform when it is not in contact with the surface, and then detects if the top platform is in contact to determine the equilibrium configurations. To solve the geometric equations of the mechanism, we use Sylvester's method of elimination. The approach obtains $4^{th}$ and $48^{th}$-degree polynomial equations for the first and second cases, respectively. Numerical examples have been applied to verify the process of analysis. The results, which are numerically calculated by software Maple, prove the validity of the analysis.
\end{abstract}

\section{INTRODUCTION}
Passive compliant parallel mechanisms are used in the area of robotics to facilitate smooth interaction with their environments, and to improve performance in manipulation tasks. This improvement can be achieved via the simultaneous regulation of force and displacement solely by the control of displacement. Therefore, the configuration of compliant mechanisms needs to be determined \cite{griffis1991kinestatic}. Compliant mechanism configuration depends on the external applied wrenches. Therefore, the inverse static analysis is crucial to revealing the system's interaction with its environment. This analysis is challenging and it involves high order non-linear equations and usually introduces multiple solutions. A reverse force analysis of a spatial three-spring system was studied in \cite{zhang1997reverse} by solving the geometric non-linear problem, and a $22^{nd}$-degree polynomial was derived. The authors proved valid solutions by the presented method. In \cite{sun1997reverse} the same solution process was applied on a different mechanism, and the authors came up with a $54^{th}$-degree polynomial. They discovered that there are only $16$ real roots among $54$. Hai-Jun used the polynomial homotopy method to present that there are as many as $70$ equilibrium configurations for a planar parallel mechanism with compliant limbs \cite{su2004inverse}. In \cite{crane2008kinematic} the authors investigated equilibrium configurations of a planner tensegrity consisting of a top and a base platform attached by one connector limb (the length of which can be controlled via a prismatic joint) and two linear springs. Their analysis showed some extraneous solutions when both springs had non-zero free lengths. A $22^{nd}$-degree univariate polynomial equation was also derived in \cite{zhang2015new} for a spatial three-spring system by a new closed-form solution.

In this paper, we present a three degree of freedom parallel planar platform mechanism. This has two components of translation and one angle of rotation. The top platform is connected with the base platform by three linear springs through revolute-joints. The top platform comes into contact with a stiff surface by a pin. This will apply a force on the top platform. We obtained equilibrium configurations of the mechanism for two cases based on the zero free length  of the springs which is commonly used in static balancing \cite{delissen2017design}. For the first case, when all of the springs' free lengths are zero, a $4^{th}$-degree polynomial equation is derived. The numerical example shows that two of the solutions are real and the other two are complex. For the second case, when the free length of one of the three springs is non-zero, the degree of the polynomial equation increases to 48. The numerical example of this case presents eight real solutions and $28$ complex ones, which satisfy the equation of force and moment. Although Sylvester's method is optimal for more complicated mechanisms like \cite{enferadi2016position,cheng2016configuration} and it should not introduce any extraneous roots, our analysis for the second case shows that there are $12$ extraneous sets that do not satisfy the equation of force and moment. For both cases, the configurations of the mechanism for the real solutions are drawn. In the first case, one of the real solutions shows that the surface is pulling on the top platform to hold the mechanism in contact with the surface, since the free lengths are zero. The second solution in the first case shows that the mechanism is in equilibrium. In case two, four real solutions are grouped together on one side of the surface while the other four real solutions are grouped on the other side.
\section{Problem Statement}
Fig. \ref{fig1} shows the mechanism that will be considered. The top platform is attached to the base platform by three springs with free lengths, $L_{01}$, $L_{02}$ and $L_{03}$, through the illustrated points. There are three coordinate systems for this mechanism: the first one is fixed, the second is coordinate system one attached to the base platform, and the third is coordinate system two attached to the top platform. The given values are:
\begin{figure}
\begin{centering}
	\includegraphics[scale=0.3]{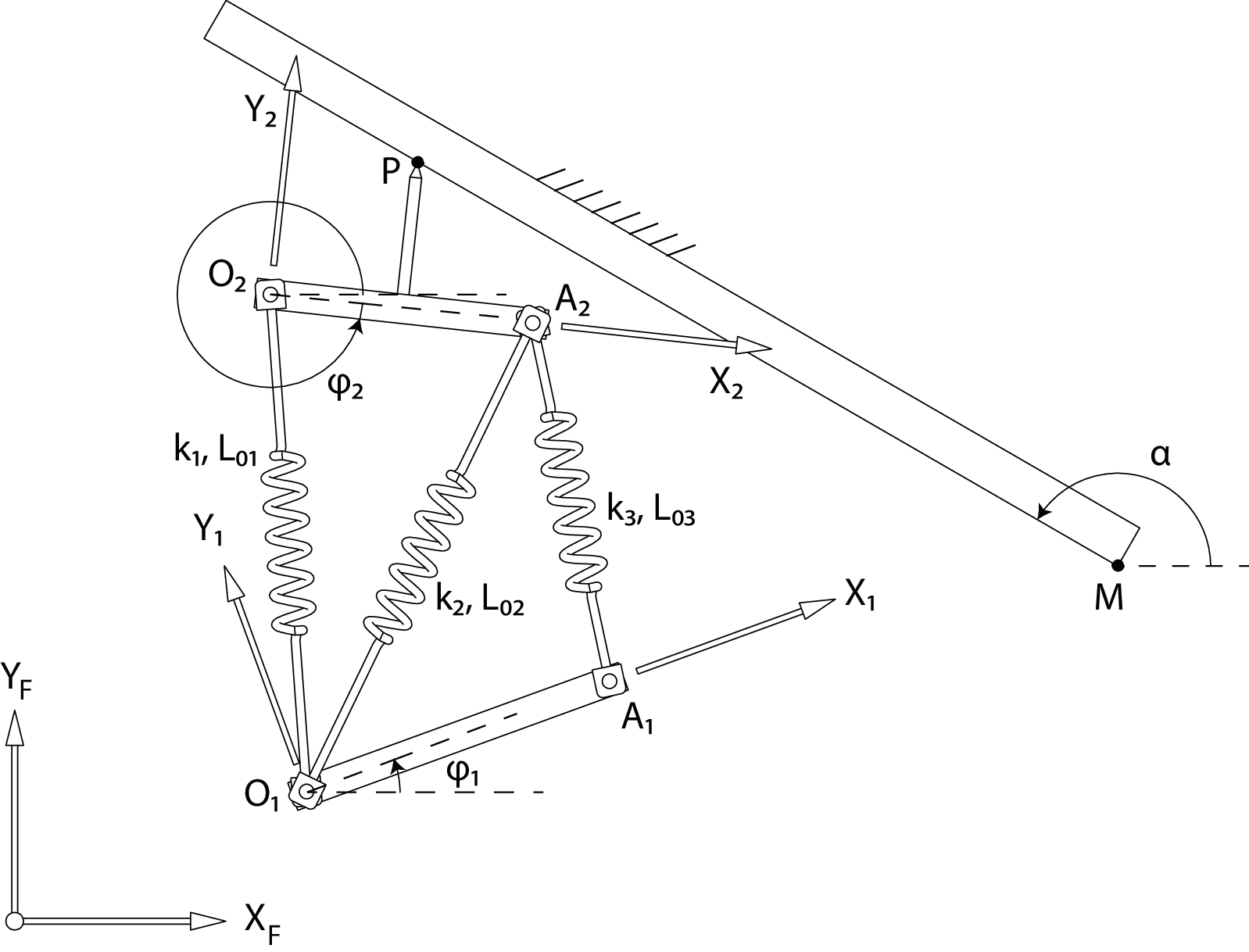}
	\par\end{centering}
\caption{The compliant planar platform mechanism\label{fig1}}

\end{figure}

\begin{itemize}
\item $^{F}\textbf{\textit{P}}_{M}$, $\alpha$, coordinates of point $M$ in the fixed coordinate
system and orientation of surface.
\end{itemize}
\begin{itemize}
\item $^{1}\textbf{\textit{P}}_{A1}$, $^{2}\textbf{\textit{P}}_{A2}$, coordinates of points $A_{1}$and $A_{2}$
in coordinate systems 1 and 2 (points are along their $X$ axes respectively).
\end{itemize}
\begin{itemize}
\item $^{2}\textbf{\textit{P}}_{P}=\left[\begin{array}{c}
p_{x2}\\
p_{y2}
\end{array}\right]$, coordinates of point $P$ in coordinate system 2.
\end{itemize}
\begin{itemize}
\item $^{F}\textbf{\textit{P}}_{O1}$, $\phi_{1}$, position and orientation of the base platform
with respect to the fixed coordinate system.
\end{itemize}
\begin{itemize}
\item $k_{i}$, $L_{0i}$, $i=1..3$, spring constants and free lengths
of the three springs.
\end{itemize}

The aim to obtain:
\begin{itemize}
\item $^{F}\textbf{\textit{P}}_{O2}$, $\phi_{2}$, position and orientation of the top platform measured
with respect to the fixed coordinate system, when the system is in equilibrium. 
\end{itemize}

\section{Solution Approach}
Sylvester dialytic method of elimination is used to solve polynomial equations in multiple variables \cite{CraneIII2008}. Furthermore, geometry of points, lines and planes have been utilized through the analysis, as well as the theory of screws, to model force and moment projection \cite{crane2009screw}.
\subsection{Calculating the Pose of the Top Platform}

The three springs are considered to be at their free lengths when the top platform is not in contact with the surface. From the corollary in \cite{nayak2018operation}, the coordinates of point $A_{2}$ can be determined in coordinate system 1 as the intersection of two circles, one centered at point $O_{1}$ and the other at point $A_{1}$. The corresponding location for point $O_{2}$ can then be determined as the intersection of a circle centered at point $O_{1}$ and a circle centered at $A_{2}$.
The equations for the circles defined by legs $O_{1}A_{2}$ and $A_{1}A_{2}$ are written as
\begin{equation}
\left(^{1}a_{2x}\right)^{2}+\left(^{1}a_{2y}\right)^{2}=L_{02}^{2},\label{eq:1}
\end{equation}
\begin{equation}
\left(^{1}a_{2x}-d_{O1A1}\right)^{2}+\left(^{1}a_{2y}\right)^{2}=L_{03}^{2}\label{eq:2}
\end{equation}
where $^{1}a_{2x}$ and $^{1}a_{2y}$ are the coordinates of point
$A_{2}$ expressed in coordinate system 1 and $d_{O1A1}$ is the distance
between points $O_{1}$ and $A_{1}$. Solving (\ref{eq:1}) for $(^{1}a_{2y})^{2}$ and substituting into (\ref{eq:2}) results in a single solution for
$^{1}a_{2x}$, as

\begin{equation}
^{1}a_{2x}=\frac{L_{02}^{2}-L_{03}^{2}+d_{O1A1}^{2}}{2d_{O1A1}}.\label{eq:3}
\end{equation}
Corresponding values for $^{1}a_{2y}$ are obtained from (\ref{eq:1})
as

\begin{equation}
^{1}a_{2y}=\pm\sqrt{L_{02}^{2}-\left(^{1}a_{2x}\right)^{2}}.\label{eq:4}
\end{equation}
The positive value for $^{1}a_{2y}$ will be selected as only this
configuration of the mechanism is of practical interest for this problem.

The coordinates of point $O_{2}$, as measured in coordinate system
1 can now be determined as the intersection of the circle centered
at point $A_{2}$, with radius $d_{O2A2}$ and the circle centered
at $O_{1}$ with radius $L_{01}$. The equations for these circles
are written as

\begin{equation}
\left(^{1}o_{2x}-{}^{1}a_{2x}\right)^{2}+\left(^{1}o_{2y}-^{1}a_{2y}\right)^{2}=d_{02A2}^{2},\label{eq:5}
\end{equation}

\begin{equation}
\left(^{1}o_{2x}\right)^{2}+\left(^{1}o_{2y}\right)^{2}=L_{01}^{2}.\label{eq:6}
\end{equation}
These two equations can be written in matrix format as

\begin{equation}
\left[\begin{array}{ccc}
	1 & A & B\\
	1 & 0 & C
\end{array}\right]\left[\begin{array}{c}
	\left(^{1}o_{2y}\right)^{2}\\
	^{1}o_{2y}\\
	1
\end{array}\right]=\left[\begin{array}{c}
	0\\
	0
\end{array}\right]\label{eq:7}
\end{equation}

\noindent where, upon substituting (\ref{eq:1}) into the expression for B,

\begin{align}
\label{eq:8}
\begin{split}
	&A=-2(^{1}a_{2y}),\\
	&B=(^{1}o_{2x})^{2}-2(^{1}o_{2x})(^{1}a_{2x})+L_{02}^{2}-d_{O2A2}^{2},\\
	&C=(^{1}o_{2x})^{2}-L_{01}^{2}.
\end{split}
\end{align}
Multiplying (\ref{eq:5}) and (\ref{eq:6}) by $^{1}o_{2y}$
yields two additional equations and the four equations can now be
written in matrix format as

\begin{equation}
\left[\begin{array}{cccc}
	0 & 1 & A & B\\
	0 & 1 & 0 & C\\
	1 & A & B & 0\\
	1 & 0 & C & 0
\end{array}\right]\left[\begin{array}{c}
	\left(^{1}o_{2y}\right)^{3}\\
	\left(^{1}o_{2y}\right)^{2}\\
	^{1}o_{2y}\\
	1
\end{array}\right]=\left[\begin{array}{c}
	0\\
	0\\
	0\\
	0
\end{array}\right].\label{eq:9}
\end{equation}
The necessary condition for finding a solution to $^{1}o_{2y}$ is that the four equations must be linearly dependent. Thus the determinant
of the coefficient matrix must equal zero, i.e.

\begin{eqnarray}
\begin{vmatrix}
	0 & 1 & A & B\\
	0 & 1 & 0 & C\\
	1 & A & B & 0\\
	1 & 0 & C & 0
\end{vmatrix}
=0.
\label{eq:10}
\end{eqnarray} 
Expansion of this determinant gives

\begin{equation}
-B^{2}+2BC-A^{2}C-C^{2}=0. \label{eq:11}
\end{equation}
Substituting (\ref{eq:8}) into (\ref{eq:11}) results in a second
order polynomial in $^{1}o_{2x}$,

\begin{equation}
D\left(^{1}o_{2x}\right)^{2}+E\left(^{1}o_{2x}\right)+F=0\label{eq:12}
\end{equation}

\noindent where

\begin{align}
\label{eq:13}
\begin{split}
	&D=-4L_{02}^{2},\\
	&E=4(^{1}a_{2x})[L_{01}^{2}+L_{02}^{2}-d_{O2A2}^{2}],\\
	&F=-[L_{01}^{2}+L_{02}^{2}-d_{O2A2}^{2}-2L_{01}(^{1}a_{2y})].
\end{split}
\end{align}
Two solutions exist for $^{1}o_{2x}$ . Corresponding values for $^{1}o_{2y}$
can be obtained by selecting the first three equations from set (\ref{eq:9})
and writing them in matrix format as

\begin{equation}
\left[\begin{array}{ccc}
	0 & 1 & A\\
	0 & 1 & 0\\
	1 & A & B
\end{array}\right]\left[\begin{array}{c}
	\left(^{1}o_{2y}\right)^{3}\\
	\left(^{1}o_{2y}\right)^{2}\\
	^{1}o_{2y}
\end{array}\right]=\left[\begin{array}{c}
	-B\\
	-C\\
	0
\end{array}\right].\label{eq:14}
\end{equation}
The coefficients $A$, $B$ and $C$ evaluate to numeric values for
each value of $^{1}o_{2x}$ and the corresponding value of $^{1}o_{2y}$
can be obtained as the third component of the vector

\begin{align}
\label{eq:15}
\begin{split}
	\left[\begin{array}{c}(^{1}o_{2y})^{3}\\
		(^{1}o_{2y})^{2}\\
		^{1}o_{2y}
	\end{array}\right] & =\left[\begin{array}{ccc}
		0 & 1 & A\\
		0 & 1 & 0\\
		1 & A & B
	\end{array}\right]^{-1}\left[\begin{array}{c}
		-B\\
		-C\\
		0
	\end{array}\right],\\
	& =\left[\begin{array}{ccc}
		-\frac{B}{A} & -\frac{A^{2}-B}{A} & 1\\
		0 & 1 & 0\\
		\frac{1}{A} & -\frac{1}{A} & 0
	\end{array}\right]\left[\begin{array}{c}
		-B\\
		-C\\
		0
	\end{array}\right].
\end{split}
\end{align}
Solving $^{1}o_{2y}$ gives

\begin{equation}
^{1}o_{2y}=\frac{C-B}{A}.\label{eq:16}
\end{equation}
\subsection{Determining if the Top Platform is in Contact}
The pose of the base platform is defined by $^{F}\textbf{\textit{P}}_{O1}$, the coordinates
of point $O_{1}$ measured in the fixed coordinate system, and the angle
$\phi_{1}$. The objective here is to determine if point $P$ is indeed
in contact with the surface based on the current pose of the base
platform. This will be accomplished by calculating the coordinates
of point $P$ in the fixed coordinate system for the current pose
of the base platform assuming that the surface did not exist and all
the springs were at their free lengths. If point $P$ lies on the
same side of the surface as the point of origin, then the top platform
is not in contact with the surface. If point $P$ lies on the opposite
side of the surface, then point $P$ must be in contact.

The coordinates of point $O_{2}$ and the angle $\phi_{2}$ have been
determined in coordinate system 1 as being when the three springs are at their
free lengths. Thus the $3\times3$ transformation matrix that transforms
planar points written in homogeneous coordinates from coordinate system
2 to coordinate system 1 can be written as

\begin{equation}
_{2}^{1}\textbf{\textit{T}}=\left[\begin{array}{ccc}
	cos\left(\phi_{2}\right) & -sin\left(\phi_{2}\right) & ^{1}o_{2x}\\
	sin\left(\phi_{2}\right) & cos\left(\phi_{2}\right) & ^{1}o_{2y}\\
	0 & 0 & 1
\end{array}\right].\label{eq:17}
\end{equation}
The transformation matrix that relates coordinate system 1 to the
fixed coordinate system can be written in terms of the current base
platform pose parameters as

\begin{equation}
_{1}^{F}\textbf{\textit{T}}=\left[\begin{array}{ccc}
	cos\left(\phi_{1}\right) & -sin\left(\phi_{2}\right) & ^{F}o_{1x}\\
	sin\left(\phi_{1}\right) & cos\left(\phi_{2}\right) & ^{F}o_{1y}\\
	0 & 0 & 1
\end{array}\right].\label{eq:18}
\end{equation}
The coordinates of point $P$ are known in coordinate system 2. The
coordinates of point $P$ in the fixed coordinate system can be calculated as

\begin{equation}
^{F}\textbf{\textit{P}}_{P}=\,_{1}^{F}\textbf{\textit{T}}\,_{2}^{1}\textbf{\textit{T}}\,^{2}\textbf{\textit{P}}_{P}.\label{eq:19}
\end{equation}
Here in (\ref{eq:19}), the coordinates of point $P$ are
written using homogeneous coordinates, where the  components of the point
are the same as the two-dimensional coordinates of that point. 

The equation of the plane which is the surface may
be written as

\begin{equation}
\textbf{\textit{r}}\cdot\textbf{\textit{S}}_{plane}+D_{0}=0\label{eq:20}
\end{equation}
where $\textbf{\textit{r}}$ is a vector from the origin to any point on the plane \cite{crane2009screw},
$\textbf{\textit{S}}_{plane}$ is the direction vector perpendicular to the plane, and
$D_{0}$ is the negative of the projection of any point on the plane
onto the direction $\textbf{\textit{S}}_{plane}$. The direction vector normal to the
plane of the surface is written as

\begin{equation}
\textbf{\textit{S}}_{plane}=\left[\begin{array}{c}
	cos\left(\alpha-\frac{\pi}{2}\right)\\
	sin\left(\alpha-\frac{\pi}{2}\right)\\
	0
\end{array}\right].\label{eq:21}
\end{equation}
The term $D_{0}$ is evaluated as

\begin{equation}
D_{0}=-\textbf{\textit{P}}_{M}\cdot\textbf{\textit{S}}_{plane}.\label{eq:22}
\end{equation}

Equation (\ref{eq:19}) is used to determine the coordinates of point
$P$ in the fixed coordinate system for the current base platform pose assuming all three springs are at their free length. The quantify $^{F}\textbf{\textit{P}}_{P}\cdot\textbf{\textit{S}}_{plane}+D_{0}$ is then evaluated. If this quantity equals
zero, then point $P$ lies in the plane of the surface. If the sign of this quantity is opposite to the sign of $D_{0}$, then point $P$ lies on the same side of the plane as the origin. If point $P$ lies on the same side of the plane as the origin, then the top platform
is not in contact with the surface. Otherwise, it is in contact.
\subsection{Finding the Equilibrium Pose when the Top Platform is in Contact} Point $E$ is defined in Fig. 2 as the intersection of the $X$ axis of coordinate system 1 with the surface. The pose of the top
platform is then defined as the distance $L$ of point $P$ from point $E$ and angle $\beta$.

\begin{figure}
\begin{centering}
	\includegraphics[scale=0.3]{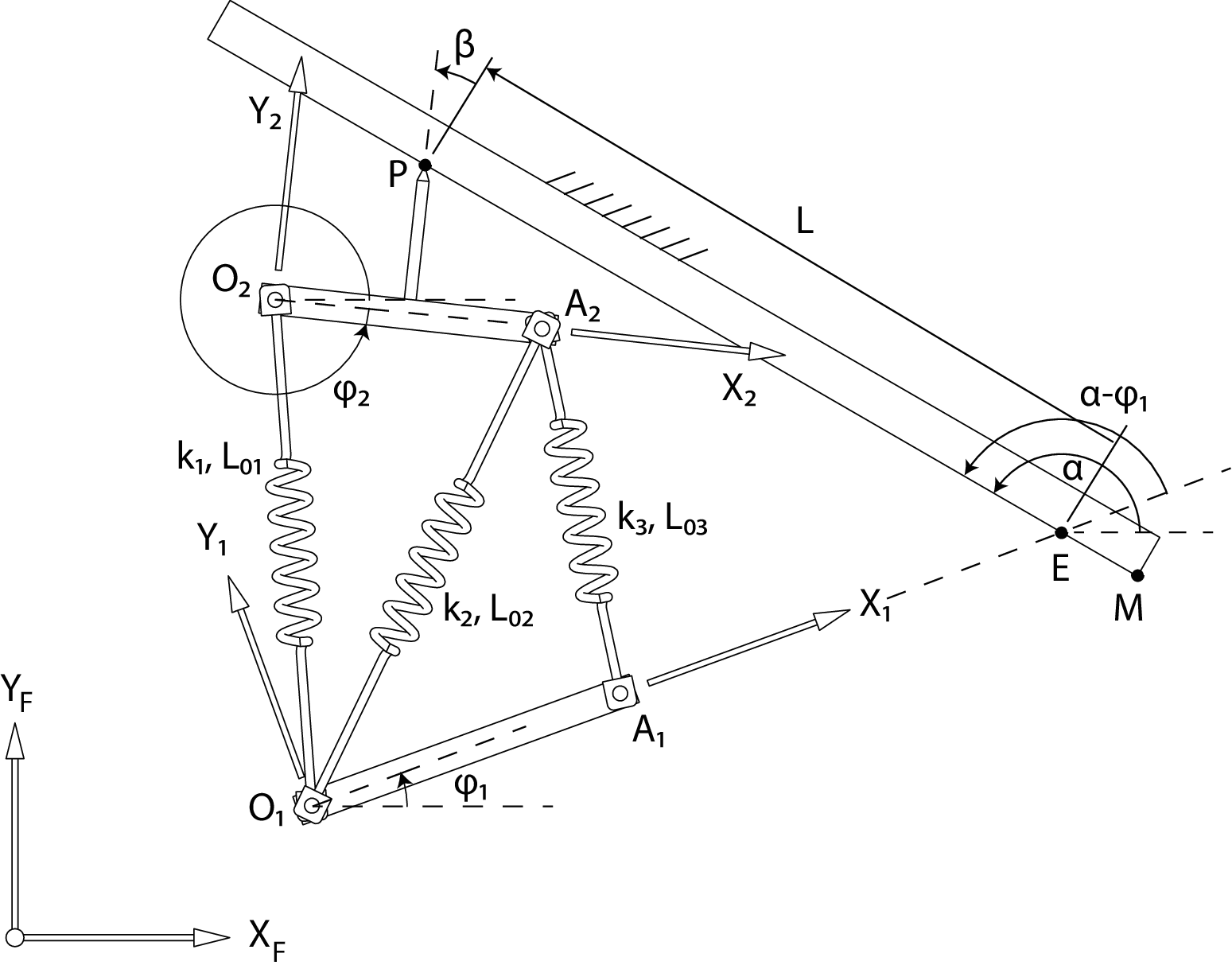}
	\par\end{centering}
\caption{Definition of point $E$, distance $L$, and angle $\beta$\label{fig:2}}

\end{figure}

Point $E$ is determined as the intersection of two lines. The first
line is defined in the fixed coordinate system by point $O_{1}$ and the angle $\phi_{1}$, and the second is defined by point $M$ and
the angle $\alpha$. The coordinates of these two lines are written
as $\left[\begin{array}{c}
\textbf{\textit{S}}_{1}\\
\textbf{\textit{S}}_{0L1}
\end{array}\right]$ and $\left[\begin{array}{c}
\textbf{\textit{S}}_{surface}\\
\textbf{\textit{S}}_{0Lsurface}
\end{array}\right]$ where

\begin{align}
\label{eq:23}
\begin{split}
	&\textbf{\textit{S}}_{1}=\left[\begin{array}{c}
		cos\left(\phi_{1}\right)\\
		sin\left(\phi_{1}\right)\\
		0
	\end{array}\right],\\
	&\textbf{\textit{S}}_{0L1}=\textbf{\textit{P}}_{O1}\times \textbf{\textit{S}}_{1}=\left[\begin{array}{c}
		0\\
		0\\
		O_{1x}sin\left(\phi_{1}\right)-O_{1y}cos\left(\phi_{1}\right)
	\end{array}\right],\\
	&\textbf{\textit{S}}_{surface}=\left[\begin{array}{c}
		cos\left(\alpha\right)\\
		sin\left(\alpha\right)\\
		0
	\end{array}\right],\\
	&\textbf{\textit{S}}_{0Lsurface}=\left[\begin{array}{c}
		0\\
		0\\
		M_{x}sin\left(\alpha\right)-M_{y}cos\left(\alpha\right)
	\end{array}\right].
	\\
\end{split}
\end{align}
The point of intersection, i.e. point $E$, can be determined from
\cite{crane2009screw} as
\begin{equation}
\label{eq:24}
\begin{split}
	&^{F}\textbf{\textit{P}}_{E}=\frac{\textbf{\textit{S}}_{surface}\times \textbf{\textit{S}}_{0Lsurface}}{1-\left(\textbf{\textit{S}}_{0L1}\cdot\textbf{\textit{S}}_{surface}\right)^{2}}\\
	&\,\,\,\,\,\,\,\,\,\,\,\,\,\,\,\,\,\,-\frac{\left(\textbf{\textit{S}}_{1}\cdot\textbf{\textit{S}}_{surface}\right)\textbf{\textit{S}}_{1}\times \textbf{\textit{S}}_{0Lsurface}}{1-\left(\textbf{\textit{S}}_{0L1}\cdot\textbf{\textit{S}}_{surface}\right)^{2}}\\
	& \,\,\,\,\,\,\,\,\,\,\,\,\,\,\,\,\,\,+\frac{\left(\textbf{\textit{S}}_{1}\times \textbf{\textit{S}}_{0L1}\cdot\textbf{\textit{S}}_{surface}\right)\textbf{\textit{S}}_{surface}}{1-\left(\textbf{\textit{S}}_{0L1}\cdot\textbf{\textit{S}}_{surface}\right)^{2}}.
\end{split}
\end{equation}

The first step of the analysis is to determine the coordinates of
points $O_{2}$ and $A_{2}$ in the fixed coordinate system, in terms
of the parameters $L$ and $\beta$. The coordinates of point $P$
are first written as

\begin{equation}
^{F}\textbf{\textit{P}}_{P}=\,^{F}\textbf{\textit{P}}_{E}+L\left[\begin{array}{c}
	cos\left(\alpha\right)\\
	sin\left(\alpha\right)
\end{array}\right].\label{eq:25}
\end{equation}
Further, the angle $\phi_{2}$ can be written as 

\begin{equation}
\phi_{2}=\alpha+\beta+\pi.\label{eq:26}
\end{equation}
The coordinates of point $P$ are given in coordinate system 2 and
written as $^{2}\textbf{\textit{P}}_{P}$ and the coordinates of points $O_{2}$ and $A_{2}$ can
now be written as

\begin{equation}
^{F}\textbf{\textit{P}}_{O2}=\,^{F}\textbf{\textit{P}}_{P}+\left[\begin{array}{cc}
	cos\left(\alpha+\beta\right) & -sin\left(\alpha+\beta\right)\\
	sin\left(\alpha+\beta\right) & cos\left(\alpha+\beta\right)
\end{array}\right]\,^{2}\textbf{\textit{P}}_{P}.\label{eq:27}
\end{equation}

\begin{equation}
^{F}\textbf{\textit{P}}_{A2}=\,^{F}\textbf{\textit{P}}_{O2}+d_{O2A2}\left[\begin{array}{c}
	cos\left(\phi_{2}\right)\\
	sin\left(\phi_{2}\right)
\end{array}\right]\label{eq:28}
\end{equation}
where $d_{O2A2}$ is the known distance between points $O_{2}$ and
$A_{2}$.

The lengths of the three springs will be referred to as $L_{1}$ $L_{2}$,
and $L_{3}$. These lengths can be expressed in terms of the coordinates
of their end points as

\begin{equation}
\left(O_{2x}-O_{1x}\right)^{2}+\left(O_{2y}-O_{1y}\right)^{2}=L_{1}^{2},\label{eq:29}
\end{equation}

\begin{equation}
\left(A_{2x}-O_{1x}\right)^{2}+\left(A_{2x}-O_{1y}\right)^{2}=L_{2}^{2},\label{eq:30}
\end{equation}

\begin{equation}
\left(A_{2x}-A_{1x}\right)^{2}+\left(A_{2y}-A_{1y}\right)^{2}=L_{3}^{2}.\label{eq:31}
\end{equation}
The terms in (\ref{eq:29}) \textendash{} (\ref{eq:31}) are the $x$
and $y$ components of the points $O_{1}$, $A_{1}$, $O_{2}$, and
$A_{2}$ written in the fixed coordinate system.

At equilibrium, the projection of the sum of the three forces in the
springs onto the direction along the surface must equal zero. The moment of these three forces with respect to point $P$ must also
equal zero. 

The directions of the lines along the springs are written as unit
vectors as

\begin{equation}
\textbf{\textit{S}}_{1}=\frac{1}{L_{1}}\left(\,^{F}\textbf{\textit{P}}_{O2}-\,^{F}\textbf{\textit{P}}_{O1}\right),\label{eq:32}
\end{equation}

\begin{equation}
\textbf{\textit{S}}_{2}=\frac{1}{L_{2}}\left(\,^{F}\textbf{\textit{P}}_{A2}-\,^{F}\textbf{\textit{P}}_{O1}\right),\label{eq:33}
\end{equation}

\begin{equation}
\textbf{\textit{S}}_{3}=\frac{1}{L_{3}}\left(\,^{F}\textbf{\textit{P}}_{A2}-\,^{F}\textbf{\textit{P}}_{A1}\right).\label{eq:34}
\end{equation}
The magnitude of the force in each spring is written as

\begin{equation}
f_{i}=k_{i}\left(L_{i}-L_{0i}\right), i=1...3.
\label{eq:35}
\end{equation}
The force projection equation that is required for equilibrium can be written as

\begin{equation}
\left(f_{1}\textbf{\textit{S}}_{1}+f_{2}\textbf{\textit{S}}_{2}+f_{3}\textbf{\textit{S}}_{3}\right).\left[\begin{array}{c}
	cos\left(\alpha\right)\\
	sin\left(\alpha\right)
\end{array}\right]=0.\label{eq:36}
\end{equation}
The moment of the three spring forces about point $P$ must also equal zero. This condition is written as
\begin{equation}
\begin{split}
	&(\,^{F}\textbf{\textit{P}}_{O1}-\,^{F}\textbf{\textit{P}}_{P})\times(f_{1}\textbf{\textit{S}}_{1})+(\,^{F}\textbf{\textit{P}}_{O1}-\,^{F}\textbf{\textit{P}}_{P})\times(f_{2}\textbf{\textit{S}}_{2})\\
	&\,\,\,\,\,\,+(\,^{F}\textbf{\textit{P}}_{A1}-\,^{F}\textbf{\textit{P}}_{P})\times(f_{3}\textbf{\textit{S}}_{3})=0.
\end{split}
\label{eq:37}
\end{equation}

The objective now is to determine the values for $L$ and $\beta$
so that (\ref{eq:36}) and (\ref{eq:37}) are satisfied.
The solution for the equilibrium pose will be conducted for two cases.
The purpose here is to show the complexity that is introduced when
the free length of a spring is a non-zero value. In each case, the
approach taken is to determine $L$ and $\beta$ so that
the projection of the sum of the three forces in the legs onto the
direction parallel to the surface is zero, and the moments of the three
forces in the legs relative to the contact point $P$ is also zero.
\subsubsection{Case i: $\mathit{L_{01}=\mathit{L_{02}}=L_{03}=0}$}

For this case, the force vectors along each spring are obtained by
substituting (\ref{eq:35}) into (\ref{eq:32}) through (\ref{eq:34})
to yield

\begin{align}
\label{eq:38}
\begin{split}
	&\mathbf{f}_{1}=f_{1}\textbf{\textit{S}}_{1}=k_{1}\left(^{F}\textbf{\textit{P}}_{O2}-\,\,^{F}P_{O1}\right),\\
	&\mathbf{f}_{2}=f_{2}\textbf{\textit{S}}_{2}=k_{2}\left(^{F}\textbf{\textit{P}}_{A2}-\,\,^{F}\textbf{\textit{P}}_{O1}\right),\\
	&\mathbf{f_{3}=\mathit{f_{3}}}\textbf{\textit{S}}_{3}=k_{3}\left(^{F}\textbf{\textit{P}}_{A2}-\,\,^{F}\textbf{\textit{P}}_{A1}\right).
\end{split}
\end{align}
Substituting  (\ref{eq:25}) into (\ref{eq:27}) and then  (\ref{eq:27}),
(\ref{eq:28}), and (\ref{eq:38}) into (\ref{eq:36}) gives

\begin{equation}
J_{1}L+J_{2}cos\left(\beta\right)+J_{3}sin\left(\beta\right)+J_{4}=0\label{eq:39}
\end{equation}
where
\begin{align}
\label{eq:40}
\begin{split}
	&J_{1}=k_{1}+k_{2}+k_{3},\\
	&J_{2}=\left(k_{1}+k_{2}+k_{3}\right)p_{x2}-d_{O2A2}\left(k_{2}+k_{3}\right),\\
	&J_{3}=-\left(k_{1}+k_{2}+k_{3}\right)p_{y2},\\
	&J_{4}=cos\left(\alpha\right)\left(\left(k_{1}+k_{2}\right)\left(E_{x}-O_{1x}\right)+k_{3}\left(E_{x}-A_{1x}\right)\right)\\ &  +sin\left(\alpha\right)\left(\left(k_{1}+k_{2}\right)\left(E_{y}-O_{1y}\right)+k_{3}\left(E_{y}-A_{1y}\right)\right).
\end{split}
\end{align}
Substituting (\ref{eq:25}) into (\ref{eq:27}) and then (\ref{eq:27})
and (\ref{eq:38}) into (\ref{eq:37}) gives
\begin{equation}
\begin{split}
&\left(K_{1}+K_{2}cos\left(\beta\right)+K_{3}sin\left(\beta\right)\right)L+K_{4}cos\left(\beta\right)\\
&+K_{5}sin\left(\beta\right)=0.\label{eq:41}
\end{split}
\end{equation}

\begin{align}
\label{eq:42}
\begin{split}
	K_{1}
	&
	=-\left(k_{1}+k_{2}+k_{3}\right)p_{x2},\\
	K_{2}
	&
	=-\left(k_{1}+k_{2}+k_{3}\right)p_{y2},\\
	K_{3}
	&
	=d_{O2A2}\left(k_{2}+k_{3}\right),\\
	K_{4}
	&
	=cos\left(\alpha\right)[p_{x2}\left(\left(k_{1}+k_{2}\right)\left(E_{y}-O_{1y}\right)+k_{3}\left(E_{y}-A_{1y}\right)\right)\\
	&+p_{y2}\left(\left(k_{1}+k_{2}\right)\left(O_{1x}-E_{x}\right)+k_{3}\left(A_{1x}-Ex\right)\right)\\
	&-d_{O2A2}\left(k_{2}\left(E_{y}-O_{1y}\right)+k_{3}\left(E_{y}-A_{1y}\right)\right)]\\
	&+sin\left(\alpha\right)[p_{x2}\left(\left(k_{1}-k_{2}\right)\left(O_{1x}-E_{x}\right)+k_{3}\left(A_{1x}-E_{x}\right)\right)\\
	&+p_{y2}\left(\left(k_{1}+k_{2}\right)\left(O_{1y}-E_{y}\right)+k_{3}\left(A_{1y}-E_{y}\right)\right)\\
	&+d_{O2A2}\left(k_{2}\left(E_{x}-O_{1x}\right)+k_{3}\left(E_{x}-A_{1x}\right)\right)],\\
	K_{5}
	& =cos\left(\alpha\right)[p_{x2}\left(\left(k_{1}+k_{2}\right)\left(O_{1x}-E_{x}\right)+k_{3}\left(A_{1x}-E_{x}\right)\right)\\
	& +p_{y2}\left(\left(k_{1}+k_{2}\right)\left(O_{1y}-E_{y}\right)+k_{3}\left(A_{1y}-E_{y}\right)\right)]\\
	& +sin\left(\alpha\right)[p_{x2}\left(\left(k_{1}-k_{2}\right)\left(O_{1y}-E_{y}\right)+k_{3}\left(A_{1y}-E_{y}\right)\right)\\
	& +p_{y2}\left(\left(k_{1}+k_{2}\right)\left(E_{x}-O_{1x}\right)+k_{3}\left(E_{y}-A_{1y}\right)\right)\\
	& +d_{O2A2}\left(k_{2}\left(E_{y}-O_{1y}\right)+k_{3}\left(E_{y}-A_{1y}\right)\right)].
\end{split}
\end{align}

The tan half-angle substitutions are now applied to (39) and (41)
\cite{CraneIII2008}. Let

\begin{equation}
x_{\beta}=tan\left(\frac{\beta}{2}\right).\label{eq:43}
\end{equation}
The sine and cosine of $\beta$ are now written as

\begin{align}
\label{eq:44}
\begin{split}
	&sin(\beta)=\frac{2x_{\beta}}{1+x_{\beta}^{2}},\\
	&cos\left(\beta\right)=\frac{1-x_{\beta}^{2}}{1+x_{\beta}^{2}}.
\end{split}
\end{align}
The expressions in (\ref{eq:44}) are substituted into (\ref{eq:39})
and (\ref{eq:41}) and these equations are then multiplied by $(1+x_{\beta})^{2}$.
The resulting equations are linear with respect to $L$, and solving
each of these for $L$ and $\beta$ and equating the results gives a fourth
order polynomial in $x_{\beta}$, which is written as

\begin{equation}
C_{4}x_{\beta}^{4}+C_{3}x_{\beta}^{3}+C_{2}x_{\beta}^{2}+C_{1}x_{\beta}+C_{0}=0\label{eq:45}
\end{equation}

\noindent where $C_{4},$ $C_{3},$ $C_{2},$ $C_{1}$ and $C_{0}$ can be determined as dependent variables on spring constants and
coordinates of $A$, $O,$ $E,$ $P$ .

Four solutions for $x_{\beta}$ can be obtained from (\ref{eq:45})
and $\beta$ is calculated as $\beta=2tan^{-1}\left(x_{\beta}\right)$
. Corresponding values for $L$ can be obtained from (\ref{eq:39})
or (\ref{eq:41}).

A numerical example is presented for this case. The given values
are:
\begin{itemize}
\item $^{F}\textbf{\textit{P}}_{M}=\left[\begin{array}{c}
19.5\\
6.25
\end{array}\right]m$, $\alpha=150.0^{\circ},$ surface parameters.
\end{itemize}
\begin{itemize}
\item $^{1}\textbf{\textit{P}}_{A1}=\left[\begin{array}{c}
5.5\\
0
\end{array}\right]m,$ $^{2}\textbf{\textit{P}}_{A2}=\left[\begin{array}{c}
4.5\\
0
\end{array}\right]m,$ coordinates of points $A_{1}$ and $A_{2}$ in coordinate system
1 and 2 (points are along $X$ axes).
\end{itemize}
\begin{itemize}
\item $^{2}\textbf{\textit{P}}_{P}=\left[\begin{array}{c}
2.25\\
2.5
\end{array}\right]m,$ coordinates of points $P$ in coordinate system 2.
\end{itemize}
\begin{itemize}
\item $^{F}\textbf{\textit{P}}_{O1}=\left[\begin{array}{c}
5\\
3.5
\end{array}\right]m,$ $\phi_{1}=20^{\circ},$ position and orientation of the base platform with respect
to the fixed coordinate system.
\end{itemize}
\begin{itemize}
\item $k_{1}=1.5$ $N/m$, $k_{2}=1.85$ $N/m,$ $k_{3}=1.45$ $N/m$, spring constants.
\end{itemize}
\begin{itemize}
\item $L_{01}=L_{02}=L_{03}=0,$  free lengths of the
springs.
\end{itemize}
Table \ref{tab:1} shows the four solutions for this case. Two of
the solutions are real and the other two are complex. All four solutions
were substituted into (\ref{eq:39}) and (\ref{eq:41}), to verify
that the equilibrium conditions are met.

\begin{figure}
\begin{centering}
	\includegraphics[scale=0.55]{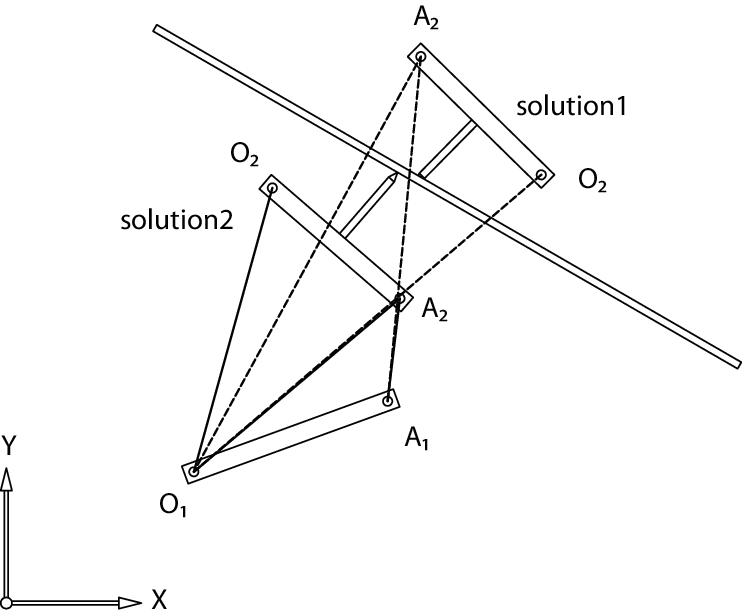}
	\par\end{centering}
\caption{Two real solutions for zero free lengths case\label{fig:3}}
\end{figure}

\begin{table}
\caption{Numerical solutions for case i) where $L_{01}=L_{02}=L_{03}=0$\label{tab:1}}
\begin{centering}
	\begin{tabular}{ccc}
		\hline 
		Solution \# & $\beta$ (radian) & $L$ (meter) \tabularnewline
		\hline 
		1 & $2.8889$ & $6.8220$\tabularnewline
		2 & $-0.1904$ & $7.3693$\tabularnewline
		3 & $-0.4294+1.8668i$ & $6.1074+8.2840i$\tabularnewline
		4 & $-0.4294-1.8668i$ & $6.1074-8.2840i$\tabularnewline
		\hline 
	\end{tabular}
	\par\end{centering}
\end{table}
Fig. \ref{fig:3} shows the two real equilibrium solutions for this numerical
example. It is interesting to note that in solution 2, the surface
is \textquoteleft pulling\textquoteright{} on point $P$, to keep that
point in contact with the surface. In solution 1, the system is in
a stable equilibrium configuration.

\subsubsection{Case ii: $\mathit{L_{01}\protect\neq0,}$ $\mathit{L_{02}}=L_{03}=0$}

This solution becomes more complicated due to the free length $L_{01}$
being non-zero. The force vectors along the three legs are obtained
by substituting (\ref{eq:35}) into (\ref{eq:32}) through  (\ref{eq:34})
as
\begin{align}
\begin{split}
	\label{eq:46}
	&\mathbf{f}_{1}=f_{1}S_{1} =k_{1}\frac{\left(L_{1}-L_{01}\right)}{L_{1}}\left(^{F}P_{O2}-\,\,^{F}P_{O1}\right),\\
	&\mathbf{f}_{2}=f_{2}S_{2}=k_{2}\left(^{F}P_{A2}-\,\,^{F}P_{O1}\right),\\
	&\mathbf{f_{3}=\mathit{f_{3}}}S_{3}=k_{3}\left(^{F}P_{A2}-\,\,^{F}P_{A1}\right).
\end{split}
\end{align}
Substituting (\ref{eq:46}) into (\ref{eq:36}) and (\ref{eq:37})
gives two equations that must be satisfied at equilibrium. The unknowns
in these equations are $L$ and $L_{1}$. (\ref{eq:29}),
when combined with (\ref{eq:25}) and (\ref{eq:27}), provides a third
equation that contains $L$, $\beta$ and $L_{1}$.
The two equations that are obtained by substituting (\ref{eq:46})
into (\ref{eq:36}) and (\ref{eq:37}) can be written as follows, upon
multiplying them throughout by $L_{1}$
\begin{equation}
\label{eq:47}
\begin{split}
	&AL_{1}=B,\\
	&CL_{1}=D
\end{split}
\end{equation}
where $A$, $B$, $C$ and $D$ are functions of $L$ and $\beta$ . Squaring
both equations and using (\ref{eq:29}) to substitute for $L_{1}^{2}$
gives two equations that can be written as
\begin{align}
\label{eq:48}
\begin{split}
	&F_{4}L^{4}+F_{3}L^{3}+F_{2}L^{2}+F_{1}L+F_{0}=0,\\
	&M_{4}L^{4}+M_{3}L^{3}+M_{2}L^{2}+M_{1}L+M_{0}=0
\end{split}
\end{align}
where the terms $F_{4}$ through $M_{0}$ are functions containing
the sines and cosines of $\beta$ as their only unknowns.
Writing (\ref{eq:48}) in matrix format gives
\begin{equation}
\left[\begin{array}{ccccc}
	F_{4} & F_{3} & F_{2} & F_{1} & F_{0}\\
	M_{4} & M_{3} & M_{2} & M_{1} & M_{0}
\end{array}\right]\left[\begin{array}{c}
	L^{4}\\
	L^{3}\\
	L^{2}\\
	L\\
	1
\end{array}\right]=\left[\begin{array}{c}
	0\\
	0
\end{array}\right].\label{eq:49}
\end{equation}
Multiplying the two equations by $L$, $L^{2}$ and $L^{3}$ gives
six additional equations. These equations, along with those in (\ref{eq:49}),
are written in matrix format as
\begin{equation}
    \begin{bmatrix}
   0 & 0 & 0 & F_{4} & F_{3} & F_{2} & F_{1} & F_{0}\\
		0 & 0 & 0 & M_{4} & M_{3} & M_{2} & M_{1} & M_{0}\\
		0 & 0 & F_{4} & F_{3} & F_{2} & F_{1} & F_{0} & 0\\
		0 & 0 & M_{4} & M_{3} & M_{2} & M_{1} & M_{0} & 0\\
		0 & F_{4} & F_{3} & F_{2} & F_{1} & F_{0} & 0 & 0\\
		0 & M_{4} & M_{3} & M_{2} & M_{1} & M_{0} & 0 & 0\\
		F_{4} & F_{3} & F_{2} & F_{1} & F_{0} & 0 & 0 & 0\\
		M_{4} & M_{3} & M_{2} & M_{1} & M_{0} & 0 & 0 & 0
  \end{bmatrix}
   \begin{bmatrix}
   		L^{7}\\
		L^{6}\\
		L^{5}\\
		L^{4}\\
		L^{3}\\
		L^{2}\\
		L\\
		1
  \end{bmatrix}
  = 
    \begin{bmatrix}
		0\\
		0\\
		0\\
		0\\
		0\\
		0\\
		0\\
		0
  \end{bmatrix}.
  \label{eq:50}
    \end{equation}
For a solution to exist, it is necessary that the eight equations be linearly dependent. Thus it is necessary that
\begin{eqnarray}
\begin{vmatrix}
	0 & 0 & 0 & F_{4} & F_{3} & F_{2} & F_{1} & F_{0}\\
	0 & 0 & 0 & M_{4} & M_{3} & M_{2} & M_{1} & M_{0}\\
	0 & 0 & F_{4} & F_{3} & F_{2} & F_{1} & F_{0} & 0\\
	0 & 0 & M_{4} & M_{3} & M_{2} & M_{1} & M_{0} & 0\\
	0 & F_{4} & F_{3} & F_{2} & F_{1} & F_{0} & 0 & 0\\
	0 & M_{4} & M_{3} & M_{2} & M_{1} & M_{0} & 0 & 0\\
	F_{4} & F_{3} & F_{2} & F_{1} & F_{0} & 0 & 0 & 0\\
	M_{4} & M_{3} & M_{2} & M_{1} & M_{0} & 0 & 0 & 0
\end{vmatrix}
=0.
\label{eq:51}
\end{eqnarray} 
Since the terms $F_{4}$ through $M_{0}$ contain the sines and cosines
of $\beta$ as their only unknown, (\ref{eq:51}) yields one equation
in the unknown $\beta$. The tan half-angle identities are substituted
into the determinant to give a univariate polynomial in the term $x_{\beta}$.
The resulting polynomial is the $48^{th}$-degree in $x_{\beta}$.

Corresponding values for $L$ for each value of $x_{\beta}$ can be
obtained by considering seven of the equations in (\ref{eq:50}).
These seven equations can be written in matrix form as

\begin{equation}
    \begin{bmatrix}
   0 & 0 & 0 & 0 & F_{4} & F_{3} & F_{2} & F_{1}\\
		0 & 0 & 0 & 0 & M_{4} & M_{3} & M_{2} & M_{1}\\
		0 & 0 & 0 & F_{4} & F_{3} & F_{2} & F_{1} & F_{0}\\
		0 & 0 & 0 & M_{4} & M_{3} & M_{2} & M_{1} & M_{0}\\
		0 & 0 & F_{4} & F_{3} & F_{2} & F_{1} & F_{0} & 0\\
		0 & 0 & M_{4} & M_{3} & M_{2} & M_{1} & M_{0} & 0\\
		F_{4} & F_{3} & F_{2} & F_{1} & F_{0} & 0 & 0 & 0
  \end{bmatrix}
   \begin{bmatrix}
   		L^{7}\\
		L^{6}\\
		L^{5}\\
		L^{4}\\
		L^{3}\\
		L^{2}\\
		L
  \end{bmatrix}
  = \\
    \begin{bmatrix}
		-F_{0}\\
		-M_{0}\\
		0\\
		0\\
		0\\
		0\\
		0
  \end{bmatrix}.
  \label{eq:50}
    \end{equation}


The coefficients $F_{4}$ through $M_{0}$ are evaluated for each
value of $x_{\beta}$ and the corresponding value for $L$ is determined
as the last term of the vector.

\begin{equation}
    \begin{bmatrix}
  L^{7}\\
		L^{6}\\
		L^{5}\\
		L^{4}\\
		L^{3}\\
		L^{2}\\
		L
  \end{bmatrix}
  =
   \begin{bmatrix}
   		0 & 0 & 0 & 0 & F_{4} & F_{3} & F_{2} & F_{1}\\
		0 & 0 & 0 & 0 & M_{4} & M_{3} & M_{2} & M_{1}\\
		0 & 0 & 0 & F_{4} & F_{3} & F_{2} & F_{1} & F_{0}\\
		0 & 0 & 0 & M_{4} & M_{3} & M_{2} & M_{1} & M_{0}\\
		0 & 0 & F_{4} & F_{3} & F_{2} & F_{1} & F_{0} & 0\\
		0 & 0 & M_{4} & M_{3} & M_{2} & M_{1} & M_{0} & 0\\
		F_{4} & F_{3} & F_{2} & F_{1} & F_{0} & 0 & 0 & 0
  \end{bmatrix}^{-1}
    \begin{bmatrix}
		-F_{0}\\
		-M_{0}\\
		0\\
		0\\
		0\\
		0\\
		0
  \end{bmatrix}.
  \label{eq:50}
    \end{equation}

A numerical example for this case will be tested as well. The given
values are the same as the previous case, except for the free length of
one of the springs, i.e. $L_{01}=1.0$.

\begin{figure}
\begin{centering}
	\includegraphics[scale=0.45]{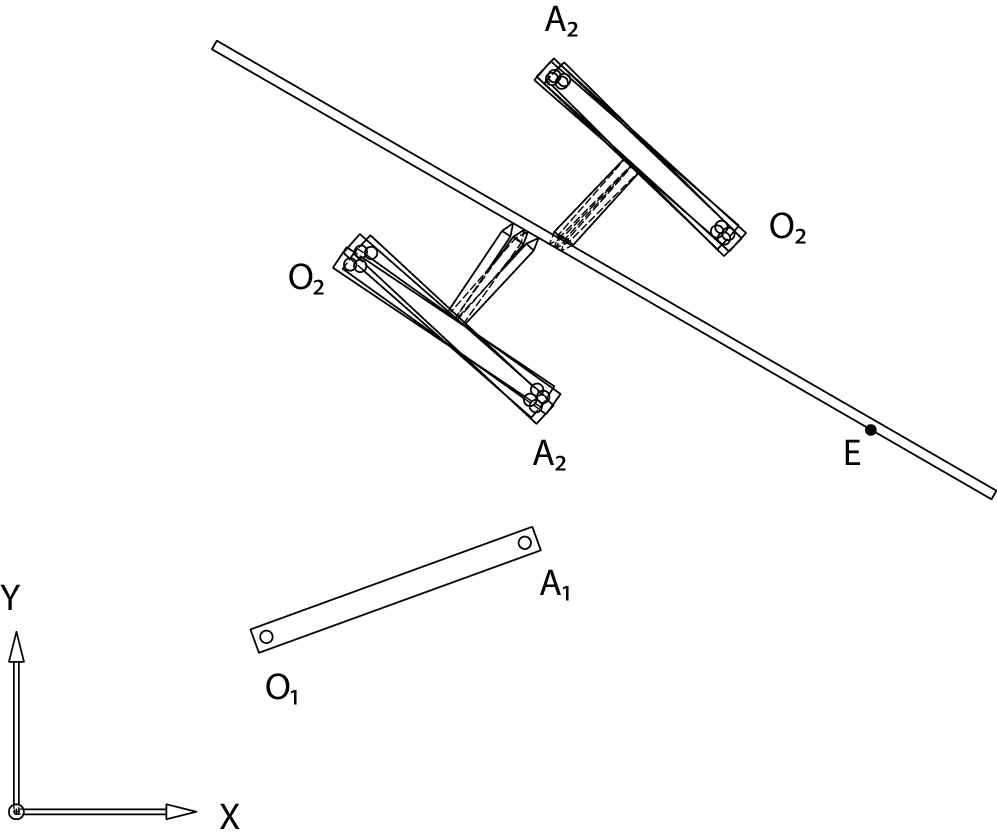}
	\par\end{centering}
\caption{Eight real solutions for one non-zero free length Case\label{fig:4}}
\end{figure}

Forty eight solutions for $L$ and $x_{\beta}$ were obtained. Each
of these solution sets were then substituted into (\ref{eq:48}) to
verify that they satisfied the conditions for equilibrium. Only $36$
of the solution sets satisfied these equations which indicates that
the solution technique introduced $12$ extraneous solutions. Table \ref{tab:2} shows the $36$ solutions, which are calculated out by software Maple, to this numeric case.
\begin{table}
\caption{Numerical solution for case ii) where $L_{01}\neq0$, $L_{02}=L_{03}=0$\label{tab:2}}
\begin{centering}
\resizebox{\columnwidth}{!}{%
	\begin{tabular}{ccc}
		\hline 
		Solution \# & $\beta$ (radian) & $L$ (meter) \tabularnewline
		\hline 
		1 & $2.9284$ & $6.8364$\tabularnewline
		2 & $2.8837$ & $6.953$\tabularnewline
		3 & $2.9468$ & $6.9906$\tabularnewline
		4 & $2.9023$ & $7.1073$\tabularnewline 
		5 & $-0.2255$ & $7.355$\tabularnewline 
		6 & $-0.1958$ & $7.6037$\tabularnewline
		7 & $-0.0970$ & $7.6834$\tabularnewline
		8 & $-0.0671$ & $7.9421$\tabularnewline
		9 & $-0.479931-1.778021i$ & $5.936438-7.867302i$\tabularnewline
		10 & $-0.479931+1.778021i$ & $5.936438+7.867302i$\tabularnewline
		11 & $-0.442435-1.882235i$ & $5.995607-7.933405i$\tabularnewline
		12 & $-0.442435+1.882235i$ & $5.995607+7.933405i$\tabularnewline
		13 & $-0.444923-1.757214i$ & $6.278938-7.749504i$\tabularnewline
		14 & $-0.444923+1.757214i$ & $6.278938+7.749504i$\tabularnewline
		15 & $-0.400698-1.867888i$ & $6.326189-7.839231i$\tabularnewline
		16 & $-0.400698+1.867888i$ & $6.326189+7.839231i$\tabularnewline
		17 & $0.148383-1.075567i$ & $7.441804-8.670424i$\tabularnewline
		18 & $0.148383+1.075567i$ & $7.441804+8.670424i$\tabularnewline
		19 & $-1.658747-1.330224i$ & $7.48042-10.277582i$\tabularnewline
		20 & $-1.658747+1.330224 i$ & $7.48042+10.277582i$\tabularnewline
		21 & $0.711855-1.023609i$ & $7.868223+0.279091i$\tabularnewline
		22 & $0.711855+1.023609i$ & $7.868223-0.279091i$\tabularnewline
		23 & $0.731613-1.712544i$ & $8.081043-9.024726i$\tabularnewline
		24 & $0.731613+1.712544i$ & $8.081043+9.024726i$\tabularnewline
		25 & $0.732104-1.71446i$ & $8.08135-9.026026i$\tabularnewline
		26 & $0.732104+1.71446i$ & $8.08135+9.026026i$\tabularnewline
		27 & $0.733525-1.712055i$ & $8.082343-9.024419i$\tabularnewline
		28 & $0.733525+1.712055i$ & $8.082343+9.024419i$\tabularnewline
		29 & $0.734018-1.713966i$ & $8.08265-9.025719i$\tabularnewline
		30 & $0.734018+1.713966i$ & $8.08265+9.025719i$\tabularnewline 
		31 & $0.768221-1.459601i$ & $8.107967-8.849827i$\tabularnewline
		32 & $0.768221+1.459601i$ & $8.107967+8.849827i$\tabularnewline
		33 & $-2.936115-1.174705i$ & $8.607065-10.526548i$\tabularnewline
		34 & $-2.936115+1.174705i$ & $8.607065+10.526548i$\tabularnewline
		35 & $1.111301-1.470257i$ & $13.473818-3.974404i$\tabularnewline
		36 & $1.111301+1.470257i$ & $13.473818+3.974404i$\tabularnewline
		\hline 
	\end{tabular}
	}
	\par\end{centering}
\end{table}
Fig. \ref{fig:4} shows the eight real equilibrium solutions for this numerical
example. It is interesting to note that for this numerical case,
there are four real solutions grouped together on one side of the
surface while the other four real solutions are grouped together on the
other side.
\section{Conclusions}
This paper presented the mathematical analysis of a Compliant planar mechanism, which comes into contact with a stiff surface, in order to find the equilibrium configurations for two cases. The analysis was presented and then investigated through numerical examples. For the first case, where all springs had zero free lengths, there were four solutions. To highlight the polynomial numerical example, two of the solutions were real and each one resulted in the top platform being located on a different side of the surface. The other two solutions were complex. The numerical example proved that all of the solutions for this case satisfy the equations of force and moment. For the second case, the free length of one of the springs was non-zero. This change affected the analysis and the solutions significantly. Although we used Sylvester's method of elimination to produce a minimal polynomial, the analysis gave a $48^{th}$-degree polynomial equation. The results of the second case, which were numerically calculated, showed that 36 solutions exist and satisfy the primary equations; eight of these 36 solutions were real and the rest were complex. The real solutions were split into two groups of four solutions, and each group resulted in the top platform being located on a different side of the surface. Future work will include an optimization for equilibrium configurations of the proposed mechanism, through a variety of methods. Two additional cases will also be considered. For the first case, two of the springs will have non-zero free lengths, and for the second all of the springs will have non-zero free lengths.
\bibliography{Parallel_Mechnism_refs.bib}


\end{document}